\documentclass[fleqn,usenatbib]{mnras}
\usepackage{newtxtext,newtxmath}
\usepackage[T1]{fontenc}
\DeclareRobustCommand{\VAN}[3]{#2}
\let\VANthebibliography\thebibliography
\def\thebibliography{\DeclareRobustCommand{\VAN}[3]{##3}\VANthebibliography}

\usepackage{graphicx}
\usepackage{amsmath}
\usepackage{hyperref}
\usepackage{multirow}
\usepackage{xcolor}

\title[The LMC Revealed in Gravitational Waves with {\it LISA}]{The Large Magellanic Cloud Revealed in Gravitational Waves with {\it LISA}}

\author[M. A. Keim et al.]{
Michael A. Keim$^{1,2}$,
Valeriya Korol$^{3,4}$,
and Elena M. Rossi$^{2}$
\\
$^{1}$Department of Astronomy, Yale University, PO Box 208101, New Haven, CT 06520-8101, USA\\
$^{2}$Leiden Observatory, Leiden University, PO Box 9513, 2300 RA Leiden, The Netherlands\\
$^{3}$Max-Planck-Institut f{\"u}r Astrophysik, Karl-Schwarzschild-Straße 1, 85741 Garching, Germany\\
$^{4}$Institute for Gravitational Wave Astronomy \& School of Physics and Astronomy, University of Birmingham, Birmingham, B15 2TT, UK 
}

\date{Received 2022 July 28; in original form 2022 July 28}
\pubyear{2022}

\begin{document}
\label{firstpage}
\pagerange{\pageref{firstpage}--\pageref{lastpage}}
\maketitle

\begin{abstract}
The {\it Laser Interferometer Space Antenna ({\it LISA})} will unveil the non-transient gravitational wave sky full of inspiralling stellar-mass compact binaries within the Local Universe. The Large Magellanic Cloud (LMC) is expected to be prominent on the {\it LISA} sky due to its proximity and its large population of double white dwarfs (DWD). Here we present the first dedicated study of the LMC with gravitational wave sources. We assemble three LMC models based on: (1) the density distribution and star formation history from optical wavelength observations, (2) a detailed hydrodynamic simulation, and (3) combining the two. Our models yield a hundred to several hundred detectable DWDs: indeed, the LMC will be a resolved galaxy in the {\it LISA} sky. Importantly, amongst these we forecast a few tens to a hundred double degenerate supernovae type Ia progenitors, a class of binaries which have never been unambiguously observed. The range in the number of detections is primarily due to differences in the LMC total stellar mass and recent star formation in our models. Our results suggest that the total number, periods, and chirp masses of {\it LISA} sources may provide independent constraints on both LMC stellar mass and recent star formation by comparing {\it LISA} observations with the models, although such constraints will be highly model-dependent. Our publicly available model populations may be used in future studies of the LMC, including its structure and contribution to {\it LISA} confusion noise.
\end{abstract}


\section{Introduction}

The milli-Hertz frequency band offers an opportunity to study persistent non-transient gravitational wave (GW) sources such as inspiralling stellar-mass compact object binaries in the Local Universe. Space-based GW observatories like the European Space Agency-led {\it Laser Interferometer Space Antenna} \citep[{\it LISA},][see also {\it TianQin} \citealt{TianQin} and {\it Taiji} \citealt{Taiji} missions]{2017arXiv170200786A} in the 2030s will detect such sources in abundance \citep[for a review see][]{LISAastroWP}. In particular, double white dwarfs (DWD) -- being the most common stellar binary remnants -- will make it possible to map our Galaxy with GWs \citep{2012PhRvD..86l4032A,2019MNRAS.483.5518K,2021MNRAS.500.4958W,geo22}. Next to the Milky Way, the Large Magellanic Cloud (LMC) will be the biggest and likely the only extended structure visible on the {\it LISA} sky \citep{2018ApJ...866L..20K, 2020A&A...638A.153K,2020ApJ...894L..15R}.

As the most massive satellite of our Milky Way, the LMC is a prime science target for {\it LISA}. Its close proximity (at $\sim$50\,kpc, \citealt{pie13,2014AJ....147..122D}) allows the detection of DWD GW signals with {\it LISA} for periods shorter than $\sim$20 minutes~\citep{2018ApJ...866L..20K}. In addition, the LMC is an active site of star formation~\citep{2009AJ....138.1243H}, which has been shown to have a critical effect on the detectable DWD population~\citep{2020A&A...638A.153K}. As a result, the LMC has been predicted to house $\sim\mathcal{O}(10^2)$ LISA-detectable DWDs, the most of any Milky Way satellite (\citealt{2020ApJ...894L..15R,2020A&A...638A.153K}; possibly including a few - several double neutron star binaries \citealt{2019MNRAS.489.4513S};~\citealt{2020ApJ...892L...9A},~\citealt{2020MNRAS.492.3061L}). GW observations will therefore extend studies of the LMC to stellar populations which are generally inaccessible with electromagnetic observatories \citep[e.g.][]{2019MNRAS.482.3656R}. Still, earlier studies have not explored the full range of observed LMC stellar masses, nor the spatial distribution of the LMC that may play a relevant role given its proximity \cite[e.g.][for double neutron stars in the Milky Way]{2022arXiv220501507S}. Including such factors and the star formation history inferred from electromagnetic observations will give a realistic picture of the LMC's DWD population and help us to understand LMC properties that LISA can explore. 

In this work we construct a realistic population of DWDs in the LMC and model their GW emission with the aim of expanding the LISA science case to the Milky Way's neighbourhood.  We estimate the number of DWDs detectable within the  nominal 4\,yr duration of the mission and we show that our results strongly depend on the assumed LMC star formation history and total stellar mass. The ultimate goal is to determine whether {\it LISA} can spatially resolve the LMC. In Section~\ref{Sec:Methods} we assemble various models of the DWD population in the LMC. These are designed to provide different total stellar masses, star formation histories, and spatial distributions of stars. We populate these models with DWDs from a fiducial binary population synthesis simulation of \citet{2020A&A...638A.153K} and we model DWDs' GW signals using the pipeline presented in \citet{kar21} to assess their detectability with {\it LISA}. In Section~\ref{Sec:Results} we describe the resulting LMC binary populations detectable by {\it LISA}, and we show how they can constrain overall LMC properties. Finally, discuss our results in Section~\ref{Sec:Discussion}. 

\section{Methods} \label{Sec:Methods} 

In this section, we first summarise LMC features known from electromagnetic radiation that are relevant for our study. Next, we construct three different LMC models: an observation-driven model (`Model~1'), a simulation-driven model (`Model~2'), and a combined model (`Model~3'). We then populate our LMC models with DWDs from a fiducial binary population synthesis simulation. Finally, we calculate the DWDs' GW signals and evaluate their detectability with {\it LISA}. 

\subsection{The LMC at electromagnetic wavelengths} \label{sec:LMCinEM}

The LMC is part of the Magellanic System consisting of the Large and Small Magellanic Clouds (SMC). It is linked to the SMC by a largely gaseous Magellanic Bridge, it is trailed by a gaseous Magellanic Stream, and extends towards the Milky Way via a gaseous Leading Arm~\citep{2005A&A...432...45B}. The origin of these features have been attributed to either strong interaction or collision between the LMC and SMC~\citep{2003MNRAS.339.1135Y, 2012MNRAS.421.2109B}, as have the LMC's subdominant arms~\citep{2016ApJ...825...20B,2019MNRAS.482L...9B}, warping of the LMC's disk~\citep{2018ApJ...866...90C, 2018ApJ...858L..21M} and central bar~\citep{2003ApJ...598L..19S,2012MNRAS.421.2109B, 2012AJ....144..106H}, the accretion of SMC stars onto the LMC~\citep{2011ApJ...737...29O,2012MNRAS.421.2109B,2020MNRAS.495...98D}, and the star formation activity of both the LMC and SMC~\citep{2003MNRAS.339.1135Y,2009AJ....138.1243H,2011A&A...535A.115I}.

A wide array of studies have been dedicated to constraining the LMC mass.
Estimates from mass-to-light ratio suggest stellar masses of 1.5$\times$10$^{9}$ M$_{\rm \odot}$ \citep{2012AJ....144....4M} to 2.7$\times$10$^{9}$ M$_{\rm \odot}$ \citep{2002AJ....124.2639V}, while rotation curve modelling suggests a neutral gas mass of 5.2$\times$10$^{8}$ M$_{\rm \odot}$ and a stellar population mass of 2.0$\times$10$^{9}$ M$_{\rm \odot}$ within a radius of 4 kpc~\citep{1998ApJ...503..674K}. The LMC's dark matter dominated total (or dynamical) mass has been estimated using kinematics of stars~\citep{1992AJ....103..447S,1998ApJ...503..674K,2014ApJ...781..121V,2020MNRAS.497.3055C,2020MNRAS.492..782W} and its interaction with surrounding galaxies and stellar streams~\citep{2013ApJ...764..161K,2016MNRAS.456L..54P,2018MNRAS.473.1218L,2018MNRAS.479..284S,2019MNRAS.487.2685E,2020MNRAS.495.2554E} with estimates arriving up to 3.1$\times$10$^{11}$ M$_{\rm \odot}$. The LMC is an active topic of research and investigations have so far ultimately relied on electromagnetic observations only. With this paper we start exploring GWs as a promising avenue for synergistic and independent studies of the LMC's mass content and star formation history, whose results would be based on completely different selection effects.

\subsection{LMC models}

\begin{figure*}
    \centering
    \quad\includegraphics[width=0.475\textwidth]{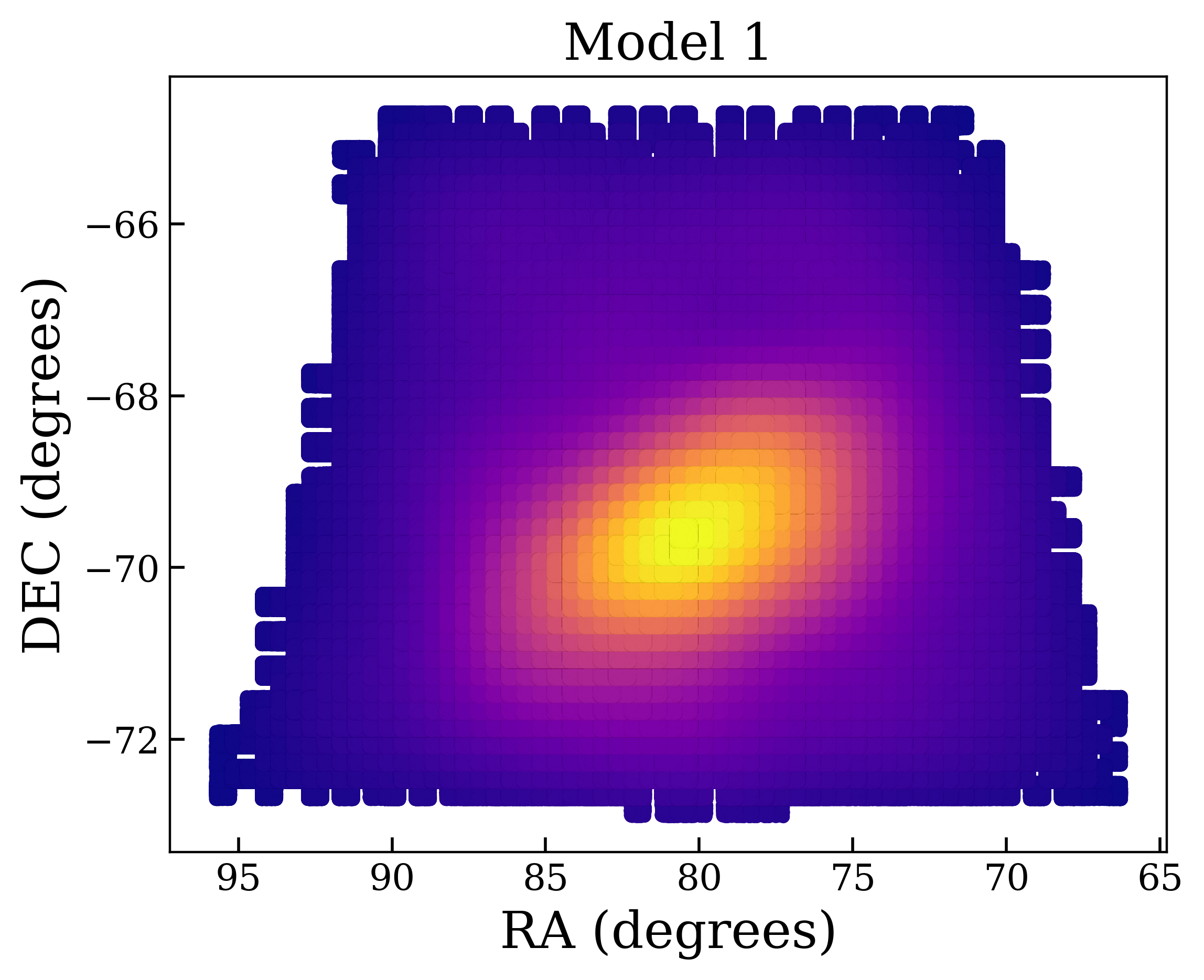}\quad
    \includegraphics[width=0.425\textwidth]{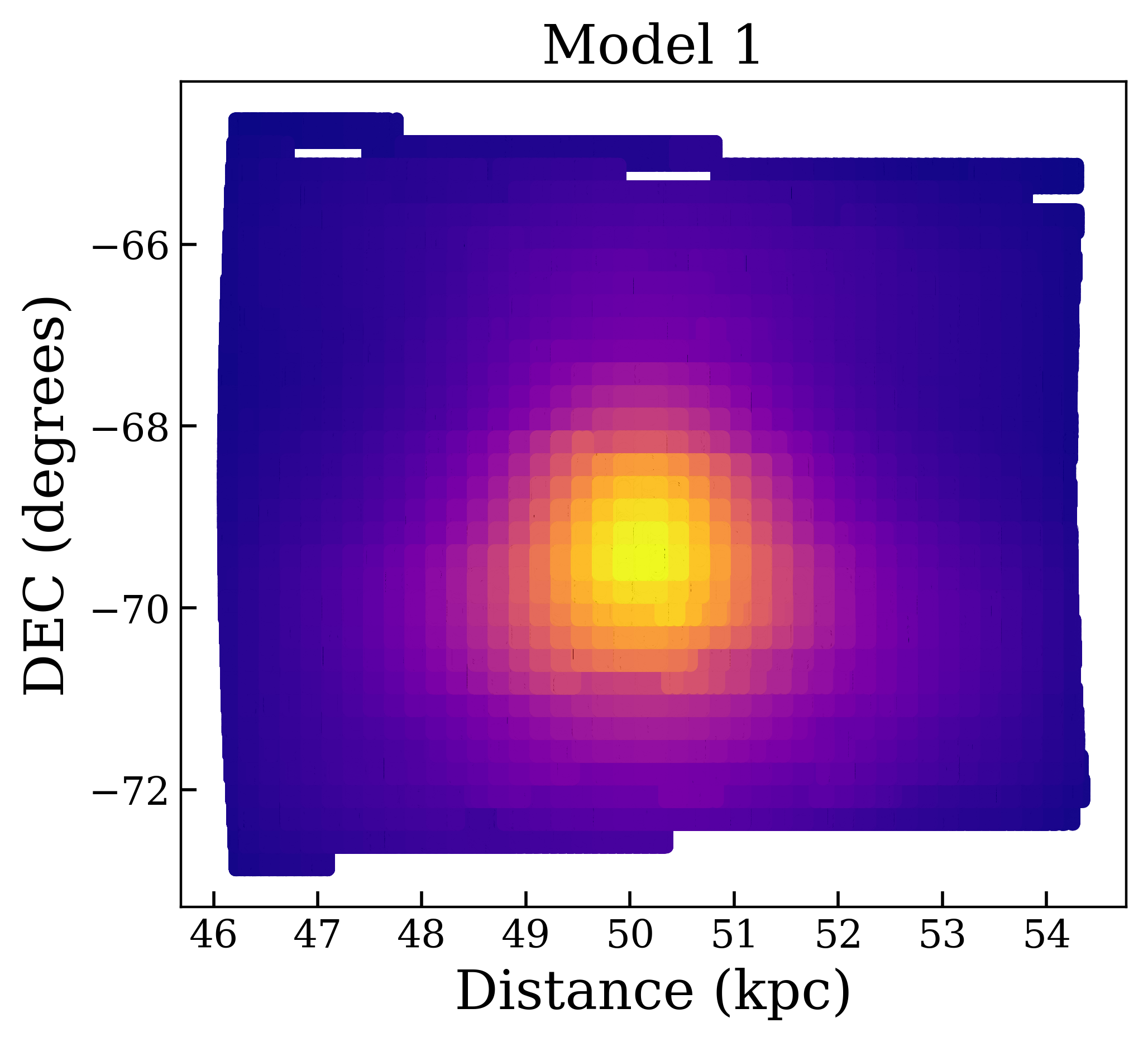}
    \includegraphics[width=0.5025\textwidth]{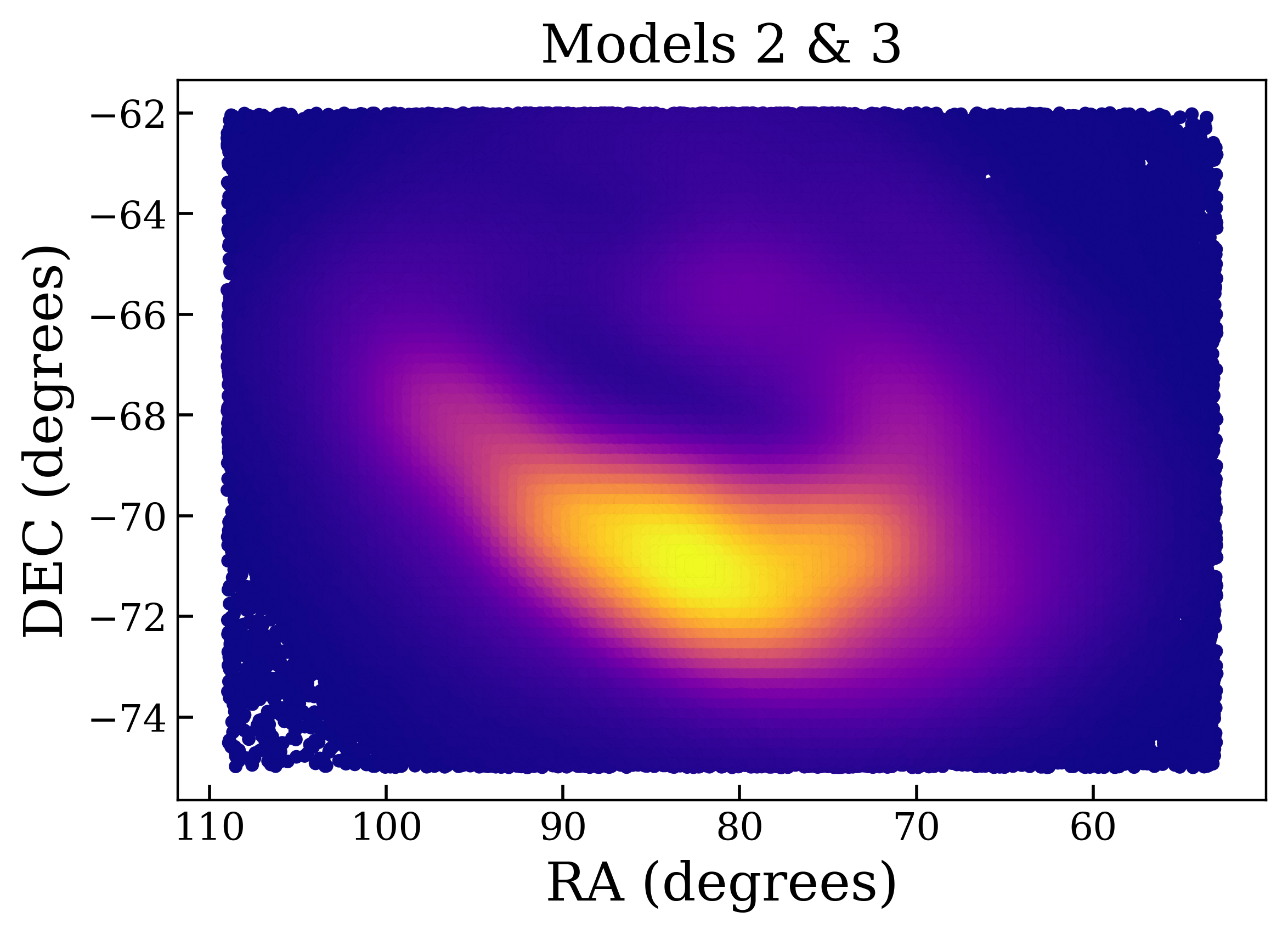}\quad
    \includegraphics[width=0.3975\textwidth]{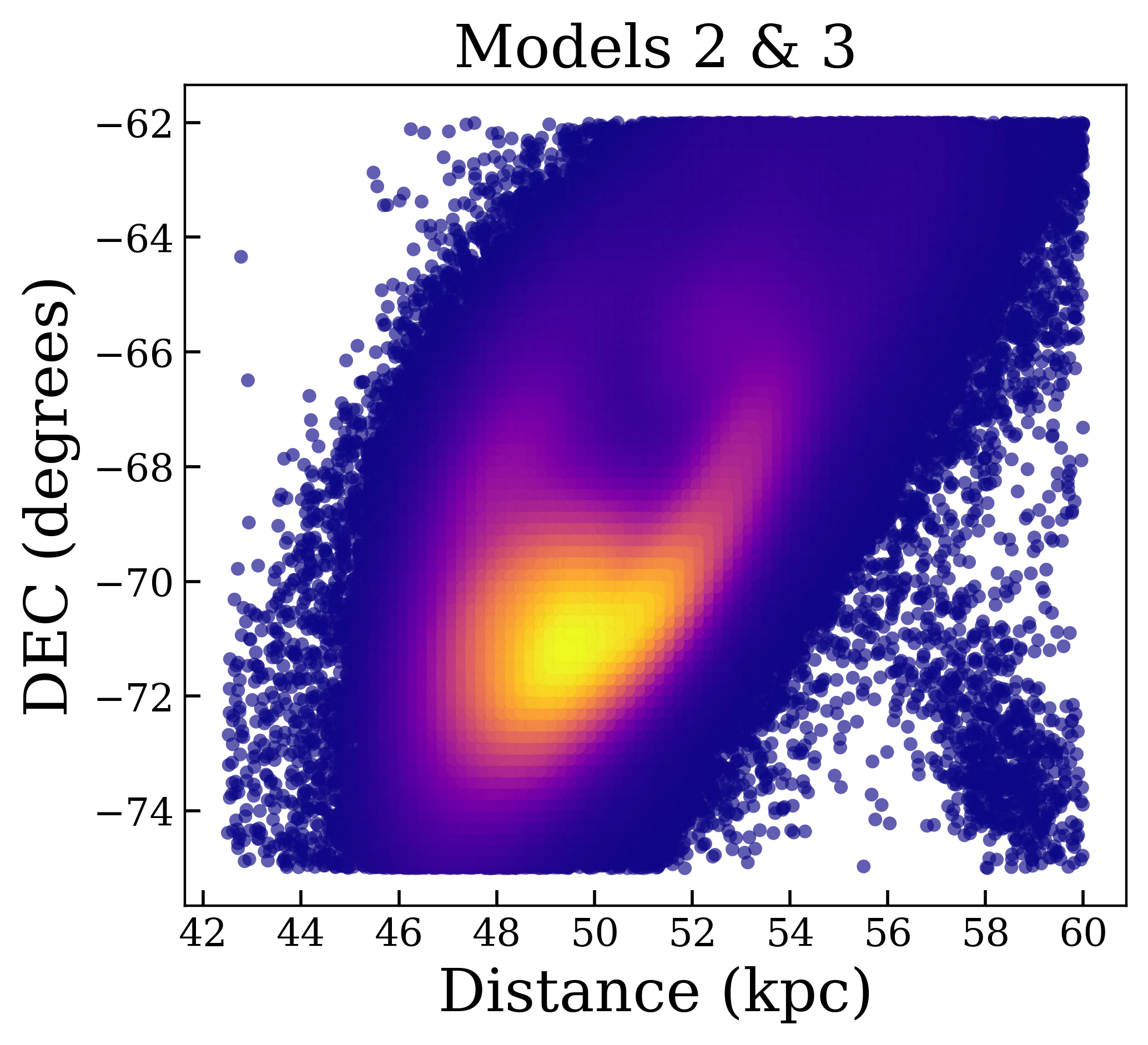}
    \caption{{\it Top panels:} Model 1 (the observation driven model)'s spatial distribution of binaries on the sky (RA, DEC; left) and along the line-of-sight (distance, DEC; right). Binaries are matched to the regions of~\citet{2009AJ....138.1243H} based on each region's observed star formation history, which are normalised so that integration gives the total mass of MSMS progenitors that should be assigned to each region. Binaries then adopt the sky location of their associated region. Distance is assigned based on a Hernquist distribution. {\it Bottom panels:} Spatial distribution of binaries on the sky (RA, DEC; left) and along the line-of-sight (distance, DEC; right) for Models 2 and 3, the simulated and combined models. Distance and sky location are assigned based on stellar particles from the simulation of~\citet{2020Natur.585..203L}. For all panels colour is indicative of local binary density with yellow corresponding to high density and blue corresponding to low density.}\label{Fig:Spatial_M1} 
\end{figure*}

{\bf Model 1: observation driven.} We first consider an observation driven model based on~\citet{2009AJ....138.1243H}'s analysis of the 20 million LMC stars observed by the Magellanic Clouds Photometric Survey~\citep{2004AJ....128.1606Z}. To derive a star formation history, the authors generate artificial colour-magnitude diagrams corresponding to stellar populations with 16 different ages between 6.3\,Myr and 15.8\,Gyr, sampled in log-space. They then fit these as a weighted sum to over a thousand different 2D sky regions of the LMC \citep[see fig. 4 of][]{2009AJ....138.1243H}. Since they apply a completeness correction in order to account for faint, unobserved stars, the resulting fit represents not only the star formation history, but also the overall mass distribution as well. 

First, for each 2D sky region, we fit the star formation history with a spline and apply a global normalisation factor so that the resulting star formation rate represents that of  main-sequence, main-sequence (MS+MS) binaries which are the progenitors of DWD binaries. Such systems have initial MS binary components with ranging from 0.14\,M$_\odot$ to 11\,M$_\odot$, though these masses will evolve subsequent to the zero age main sequence due to stellar processes and binary interactions resulting in mass loss and mass transfer. Analytically, this normalisation constant $c$ is obtained by solving
\begin{equation}\label{Eq:GME}
    c \sum_{i=1}^{1376} \int_{t_0 \: = \: 10 \: \textrm{Gyr}}^{t_{\rm f} \: = \: 0 \: \textrm{Gyr}} \textrm{SFR}_ i(t) dt = \sum_{j=1}^{N} M_{{\rm MS+MS}, j},
\end{equation}
where 1376 is the total number of regions in \citet{2009AJ....138.1243H}, SFR$_{i}$ is the region $i$'s spline-fit star formation rate, $M_{{\rm MS+MS}, j}$ is the initial mass of each MS+MS progenitor binary, and $N$ is the total number of binaries. Given that most LMC stars were born in the last 10\,Gyr with earlier times being largely quiescent~\citep[e.g.][]{1977ApJ...216..372B,2009AJ....138.1243H,2019svmc.confE..20R}, on the left hand side of Eq.~\eqref{Eq:GME} we integrate in lookback time from $t_0$ = 10\,Gyr until the present day $t_{\rm f}=0$\,Gyr. 
The density distribution of binaries across the LMC regions is then determined based on each 2D sky region's total mass in binaries,
\begin{equation}
M_{i} = \int_{t_0 \: = \: 10 \: \textrm{Gyr}}^{t_{\rm f} \: = \: 0 \: \textrm{Gyr}} c \, \textrm{SFR}_{i}(t) \, dt.    
\label{eq:Mi}
\end{equation} 
We assign formation times by dividing the total masses of binaries by the relevant region's spline-fit star formation rate and advancing in time by $\Delta t = M_{{\rm MS+MS},j}/$SFR$_{i}(t)$.  To account for periods of quiescence in a region which would make $\Delta t$ an over estimate, if $\Delta t$ exceeds 100 Myr we instead evaluate the formation time over finite time steps $dt$, continually recalculating $dm_k = $~SFR$_{i}(t_k) \times dt$ until reaching $\sum dm_i = M_{{\rm MS+MS},j}$.

Since the mass distribution found by integrating LMC regions only accounts for 2D sky distribution, to find the distance from the Sun we adopt a Hernquist potential \citep{1990ApJ...356..359H}. We consider the transform \citep[e.g.][]{2001ApJ...548..712W}
\begin{align}
	x & = - \cos(\delta)\sin(\alpha-\alpha_{\rm LMC}) d \label{Eq:Transform} \\
	y & = \left[ \sin(\delta)\cos(\delta_{\rm LMC}) - \sin(\delta_{\rm LMC})\cos(\delta) \cos(\alpha-\alpha_{\rm LMC}) \right] d \notag \\
	z & = d_{\rm LMC} - \left[ \sin(\delta)\sin(\delta_{\rm LMC}) +  \cos(\delta)\cos(\delta_{\rm LMC})\cos(\alpha-\alpha_{\rm LMC}) \right] d \notag
\end{align}
to convert right ascension (RA) $\alpha$, declination (DEC) $\delta$, distance $d$ coordinates to the $(x, y, z)$ frame for the LMC, where $(\alpha_{\rm LMC}, \delta_{\rm LMC}, d_{\rm LMC})$ is set to ($80^\circ, -69^\circ, 50\,\text{kpc})$. The distance to the equatorial plane ($z=0$) for each LMC region is then given by
\begin{equation}
    d_{z=0} = \frac{d_{\rm LMC}}{\sin(\delta)\sin(\delta_{\rm LMC}) + \cos(\delta_{\rm LMC})\cos(\delta)\cos(\alpha-\alpha_{\rm LMC})}.
    \label{Eq:dflat}
\end{equation}
For a region with a given (RA, DEC), Eqs.~(\ref{Eq:Transform}) and (\ref{Eq:dflat}) give the corresponding $(x, y)$ coordinates of that region's binaries. Binaries are then assigned a $z$ value in the interval $[-4,4]$\,kpc using a Hernquist profile, taking the mass density as a probability density so that
\begin{equation}
    P(z) \propto 
    \begin{cases}
    r^{-1}(z) \times (r(z) + r_{\rm hern})^{-3} & \text{for } r(z) > r_{\rm min}, \\
    r_{\rm min}^{-1} \times (r_{\rm min} + r_{\rm hern})^{-3} & \text{for } r(z) \leq r_{\rm min},
    \end{cases}
    \label{Eq:Hernquist}
\end{equation}
where $r_{\rm min}=0.1$\,kpc, $r_{\rm hern}$ is taken as 2\,kpc to reflect the full-width-at-half-maximum of the distribution of Cepheid variable stars~\citep{2016AcA....66..149J}, and $r(z)=\sqrt{x^2 + y^2 + (z)^2}$. Finally, we invert Eq.~(\ref{Eq:Transform}) to find a binary's final, deprojected coordinates. 

The distribution of binaries on the sky based on 2D sky regions, as well as the distribution along the line of sight following Eq.~(\ref{Eq:Hernquist}), is given in top panels of Fig.~\ref{Fig:Spatial_M1}. We note that this simplistic spatial distribution lacks small scale LMC features (cf. Section~\ref{sec:LMCinEM}). In addition, the symmetric $z$-axis limits our examination of how {\it LISA} may utilise DWD distance to probe the LMC structure, especially crucial as lack of dust extinction makes GW ideal for such a purpose. It is therefore useful to consider a model with a more sophisticated spatial distribution as explored in the following section.


{\bf Model 2: simulation driven}. To obtain a more realistic spatial distribution that includes characteristic LMC structural features and a realistic stellar distribution along $z$-axis, we utilise the N-body hydrodynamical simulation of \citet{2020Natur.585..203L}. It models the entire Magellanic system including both LMC and SMC, the Magellanic Bridge, the Magellanic Stream, and the Leading Arm.  
This simulation consists of $\sim$6.5$\times$10$^5$ particles; their spatial distribution  projected on the sky is shown in Fig.~\ref{Fig:Spatial_M1} (bottom panels). We highlight that the simulation clearly features a northern arm, and its central bar has the same characteristic warp as discovered by \citealt{2003ApJ...598L..19S}. While this model does not perfectly replicate the LMC's structure -- e.g. having a northern arm that is closer to Earth than the central bar~\citep{2016AcA....66..149J} -- it nevertheless provides a spatial distribution with detailed realistic features that potentially can be probed with GW observations. 

We populate the model LMC by assigning DWD binaries the ages and locations of the simulated stellar particles. We note that the bulk of simulated particles are formed 6\,Gyr ago, which represents the starting point of \citet{2020Natur.585..203L}'s simulation. For DWD in these older particles we randomly assign a formation time between 0 and 4\,Gyr. In this way we obtain an alternative star formation history from that of the observation driven model.



{\bf Model 3: combined.} For a final, `combined' model, we use the best features of the previous two models: the data-driven star formation history of Model 1 and the spatial distribution from Model 2. To achieve this in practice, we match each particle simulation particle of \citet{2020Natur.585..203L}'s simulation to the nearest 2D sky region in Model 1 \citep[cf. fig. 4 of][]{2009AJ....138.1243H} and adopt its star formation history. In this way, we keep the detailed structure of the LMC seen in bottom panels of Fig.~\ref{Fig:Spatial_M1}, but end up with a temporal distribution that is in agreement with electromagnetic observations.

\subsection{DWD synthetic population} \label{Sec:mockDWD}

We employ a synthetic collection of DWDs from \citet{2020A&A...638A.153K} generated using the {\sc SeBa} stellar and binary evolution module \citep{1996A&A...309..179P,2001A&A...365..491N,2012A&A...546A..70T}. Amongst their models we use the one with a fixed binary fraction of 50\,per cent and a fixed metallicity $Z=0.001$. Although LMC stars present a range of metallicities, \citet{2020A&A...638A.153K} showed that this assumption has only a moderate effect on the number of {\it LISA} DWD detections in Milky Way satellites. Note that by using a fixed binary fraction we neglect a potential correlation between the metallicity and primordial binary fraction that has been found in the Solar neighbourhood. \citep{2018ApJ...854..147B,2019MNRAS.482L.139E,2019ApJ...875...61M}. 

As the next step of our modelling, we adapt binaries' parameters (as provided by {\sc SeBa}) to resemble LMC's star formation history and spatial distribution. Specifically, we use the DWD formation time, i.e. the time it takes to turn both main sequence stars in the binary into white dwarfs, the orbital separation at DWD formation, and the white dwarf masses. First, to account for the LMC stellar mass of $\sim\mathcal{O}(10^9)$\,M$_{\odot}$, we scale up our DWD population. We estimate the number of binaries injected in each LMC model based on the total stellar mass $M_\star$ by linearly re-scaling the synthetic collection by
\begin{equation}
    N_{\rm DWD} = \frac{M_\star}{M_{\rm {\sc SeBa}}} N_{\rm DWD,{\sc SeBa}},
\end{equation}
where $N_{\rm DWD,{\sc SeBa}}$ is the number of DWDs in the synthetic population and $M_{ \rm {\sc SeBa}}$ is the total simulated population mass. We consider several LMC mass estimates in the range between $1.5\times10^9$\,M$_\odot$ and $3.0\times10^9$\,M$_\odot$~\citep{1998ApJ...503..674K,2002AJ....124.2639V,2012AJ....144....4M}. Second, we employ our three alternative LMC models to assign sky locations and distances to synthetic DWDs. Third, we assign DWD ages according to the models' star formation history. Lastly, we adjust binaries' orbital separation to reflect energy loss from GW emission since the binaries first became DWD until the present day. This step allows us to determine which binaries are emitting in the {\it LISA} band today, and to exclude those that have already merged. We compute DWD present-day GW frequencies as
\begin{equation}\label{Eq:EvolvedF}
    f(t) = \frac{3^{3/8}}{2^{9/8} \pi} \left(\frac{96}{5}\right)^{-3/8} \left( \frac{G {\cal M}}{c^3}\right)^{-5/8} (\tau_{\rm merge} - t)^{-3/8},
\end{equation}
where $G$ is the gravitational constant, $c$ is the speed of light, the chirp mass ${\cal M} = (m_1 m_2)^{3/5}/(m_1 + m_2)^{1/5}$ with $m_1$ and $m_2$ being the primary and the secondary white dwarf masses, and $\tau_{\rm merge}$ is the merger time calculated from the initial frequency at the time of DWD formation \citep[e.g.][]{Maggiore}. 

\begin{figure}
    \centering
    \includegraphics[width=\columnwidth]{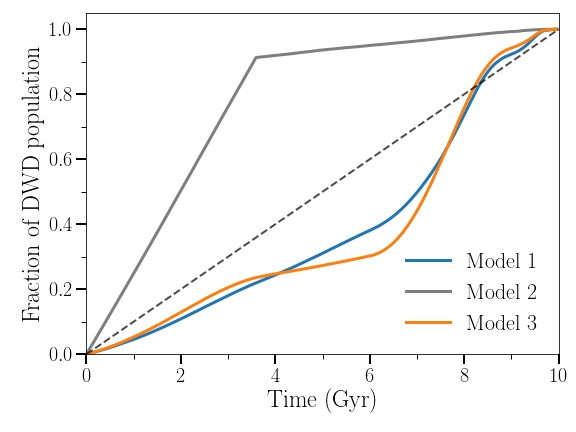}
    \caption{The DWD formation history for Model 1 (observation driven) in blue, Model 2 (simulation driven) in grey, and Model 3 (combined) in orange. The $y-$axis represents the cumulative fraction of the total DWD population emitting in the {\it LISA} frequency band today as they formed over time in the LMC ($x$-axis). As a reference, we also show the cumulative population for a constant star formation history as dashed black line (see Fig.~\ref{Fig:SFR_eff27} for the historic star formation \textit{rate}).}\label{Fig:SFR_eff} 
\end{figure}

In Fig.~\ref{Fig:SFR_eff} we summarise the obtained DWD formation histories in our three LMC models: observation driven (Model 1 in blue), simulation driven (Model 2 in grey) and combined (Model 3 in orange). We highlight a stark difference between the observation driven (Model 1) and simulation driven (Model 2) models. In the later, DWD formation is very steep up to $\sim4$\,Gyr and negligible since. In the former, the star formation rate trend is reversed: DWD formation is shallow up to 6\,Gyr followed by a significant grows in the last 3\,Gyr. In particular, a burst around 500-800 Myr ago is responsible for a much larger young binary population. This is best seen with logarithmic spacing as given in Fig.~\ref{Fig:SFR_eff27}, and, along with a more recent burst $\sim$100 Myr ago is thought to correspond with interactions with the SMC \citep{1996MNRAS.278..191G,2000AcA....50..355P,2003MNRAS.339.1135Y,2009AJ....138.1243H,2011A&A...535A.115I,2012MNRAS.421.2109B}. More quantitatively, we read from Fig.~\ref{Fig:SFR_eff} that in Model 2 the DWD population reaches 90\, per cent already at $\sim4$\,Gyr, while in Model 1 and 3, which share the same observation driven star formation history, it reaches 90\, per cent only at 9\,Gyr. This difference makes a significant impact on the size and the properties of the {\it LISA} detectable DWDs. As discussed in \citet{2020A&A...638A.153K}, recent star formation activity adds newly more massive short period binaries to the {\it LISA} band that are also more easy to detect; in old stellar populations more massive short period binaries have already merged, while long period binaries have not yet evolved into the {\it LISA} band. We also note a slight difference in the DWD formation histories of the combined model and observation driven model. These arise because the number of binaries falling into the different LMC 2D sky regions is based on a different underlying mass distribution. Thus, the various star formation pixels in Models 1 and 3 have a different weight in the overall star formation history.

\subsection{DWD detectability with {\it LISA}} \label{Sec:DA}

As the result of the methodology described so far, we obtain a catalogue of $\sim\mathcal{O}(10^6)$ LMC DWDs emitting in the {\it LISA} frequency band. As the next step, we use the {\it LISA} data analysis pipeline presented in \citet{kar21}, based on an signal-to-noise (S/N) evaluation using an iterative scheme for the estimate of the confusion foreground generated by Milky Way's GW sources \citep[see also][]{tim06,cro07,nis12}. We use {\it LISA}'s instrumental noise requirements defined in a technical note by \citet{LISAdoc}.
We set a {\it LISA} mission duration of 4\,yr and an S/N detection threshold of 7.

The scheme begins with the generation of the signal measured by {\it LISA}, by computing quasi-monochromatic wave-form for each DWD and by projecting it on the {\it LISA} arms for a given duration of the mission. Such a wave-form can be fully determined by 8 parameters: amplitude ${\cal A}$, frequency $f$, frequency derivative or chirp $\dot{f}$, sky position $(\theta,\phi)$ in ecliptic heliocentric coordinates, orbital inclination $\iota$, polarisation angle $\psi$ and initial orbital phase $\phi_0$ \citep[e.g.][]{cut98}. The first five parameters represent the outcome of our modelling procedure (cf. Section~\ref{Sec:mockDWD}), while the remaining three are angular parameters that we assign randomly: $\iota$ is sampled from a uniform distribution in $\cos \iota$, while $\psi$ and $\phi_0$ are sampled from a flat distribution.
We compute the gravitational amplitude as 
\begin{equation} \label{eqn:amp}
    {\cal A} = \frac{2(G{\cal M})^{5/3}(\pi f)^{2/3}}{c^4 d}, 
\end{equation}
and the frequency derivative as
\begin{equation}\label{eqn:fdot}
\dot{f}=\frac{96}{5} \pi^{8/3} \left( \frac{G{\cal M}}{c^3} \right)^{5/3} f^{11/3}.
\end{equation}

For each binary in the catalogue, the pipeline then computes S/N and performs an Fisher information matrix (FIM) analysis in order to estimate the accuracy of the parameter recovery. We recall that FIM for each recovered source can be evaluated as 
\begin{equation}
	\left. F_{ij} = \left( \frac{\partial h(\vec{\theta})}{\partial \theta_i} \bigg| \frac{\partial h(\vec{\theta})}{\partial \theta_j}  \right) \right\vert_{\vec{\theta} = \vec{\theta}_\mathrm{true}},
	\label{eq:fim}
\end{equation}
where $h$ is the template of each signal, and $\vec{\theta}$ represents the wave-form parameter vector. The inverse of $F_{ij}$ yields the covariance matrix with the diagonal elements being the mean square errors on each parameter $\theta_i$, and the off-diagonal elements describing the correlations between the parameters $\rho_{\theta_i \theta_j}$. We remark that errors derived via FIM analysis are valid for relatively high S/N \citep[e.g.][]{cut98}. Thus, we expect that in some cases the errors may be underestimated. A full Bayesian parameter estimation is required to derive more realistic uncertainties \citep[e.g.][]{bus19,katz22,fin22}.


\section{Results} \label{Sec:Results}

\subsection{Number of detections}
\begin{figure}
    \centering
    \includegraphics[width=0.45\textwidth]{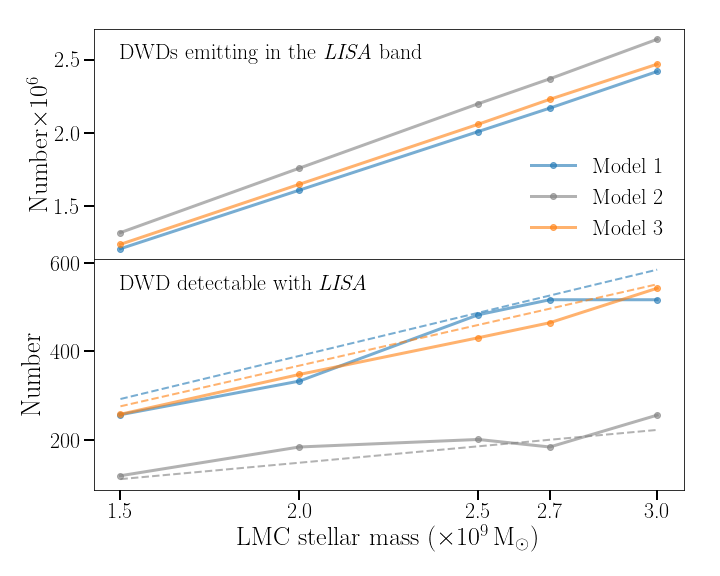}\quad
    \includegraphics[width=0.45\textwidth]{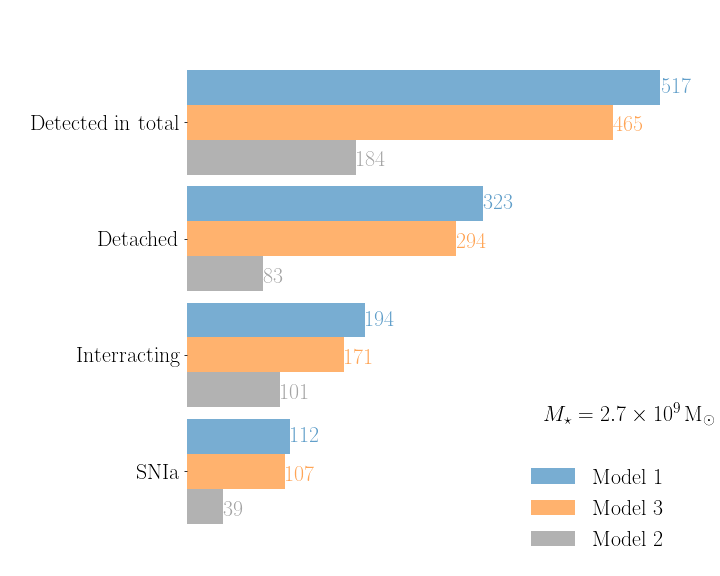}
    \caption{Summary plots for the number of detected LMC DWDs. {\it Top panels}: solid lines represent the trend for the total DWD number emitting in the {\it LISA} band and those detectable with {\it LISA} within 4\,yr of mission; dashed lines show linear regression fit (cf. Eq.\ref{Eq:Regression}). {\it Bottom panel}: the break down of detected DWDs in different categories for a fixed LMC mass of $2.7\times10^{9}$\,M$_\odot$.}\label{Fig:results} 
\end{figure}
Out of $\sim\mathcal{O}(10^6)$ emitting in the {\it LISA} frequency band, we find from a few to several $\sim\mathcal{O}(10^2)$ to be detectable with S/N$>7$ assuming 4\,yr for the mission duration with 100\,per cent duty cycle \citep[however see][]{LISAmissionduration}. Our results are reported in Table~\ref{Table:Mass}. All detected LMC DWDs have $f>1.7\,$mHz (orbital period $<20\,$min), which reflects {\it LISA}'s selection effects with frequency and distance \citep[e.g. see fig.~1.1 of][]{LISAastroWP}. Fig.~\ref{Fig:results} illustrates our results. In the top panel it shows the trend for the total number of DWDs emitting in the {\it LISA} band and the number of detectable ones with the LMC mass; in the bottom panel it shows the break down of detectable DWDs in different categories (detached, interacting and supernova type Ia progenitors, SNIa) for a fixed LMC mass of $2.7\times10^{9}$\,M$_\odot$. We find that the number of DWDs emitting in the {\it LISA} band primarily scales with the LMC stellar mass, with a little difference between our LMC models. The number of detected DWDs across considered LMC models ranges between 119 and 543 (see Table~\ref{Table:Mass}). As we keep DWD synthetic population fixed, differences in the number of detections can be attributed to differences in the LMC stellar mass and/or star formation history and/or spatial distribution within the LMC. We next discuss the relative importance of these model ingredients. 

It is immediately evident from Fig.~\ref{Fig:results} that the number of DWDs (total in the {\it LISA} band and detected) scale linearly with the assumed LMC stellar mass. For a fixed LMC mass, by comparing Model 1 and 3 (i.e. models with the same star formation history but different spatial distributions), we can deduce that the spatial distribution alone has a minor impact of less than 10\,per cent. This is because the LMC's size is still fairly small compared to its distance so that re-distribution does not cause substantial differences. The largest variation (up to $\sim 60$\,per cent) in total number of detected binaries is due to different underlying star formation history. This is evident by comparing Model 2 and Model 3. 

Our results reveal that up to 50-200 {\it LISA} detectable DWDs may be in an interacting phase, with possible electromagnetic counterpart signals. Here we flag as `interacting' those binaries in which one of the two white dwarfs overfill its Roche Lobe as defined by~\citep{1983ApJ...268..368E}
\begin{equation} \label{Eq:Roche}
    r_{\rm RL} = \frac{0.49 q^{2/3}a}{0.6 q^{2/3} + \log{(1 + q^{1/3})} },
\end{equation} 
where $q=m_2/m_1$ is the mass ratio and $a$ is the DWD orbital separation derived from Eq.~\eqref{Eq:EvolvedF} using the Kepler's law.
Material extending beyond this tear drop shaped gravitational equipotential is not bound to its white dwarf and mass transfer with the binary partner may occur, potentially impacting the GW signal. From this point onward, it is yet unclear if the system onsets a stable mass transfer or merges \citep[e.g.][and \citet{LISAastroWP} for a review]{mar04,Shen15,tau18}. Thus, we do not attempt to model further the evolution of these systems in the present work. 

Importantly, amongst detected DWDs 25-125 (including both detached and interacting) have chirp masses $>0.6$\,M$_{\rm \odot}$. This chirp mass threshold corresponds to an equal mass binary with a total mass exceeding the Chandrasekhar limit. When such binaries merge, they may result in SNIa explosions \citep[e.g.][]{1984ApJ...277..355W}, although an alternative outcome may be an accretion-induced collapse resulting in a neutron star \citep[e.g.][]{1985ApJ...297..531N,2012ApJ...748...35S}.

\subsection{Estimated {\it LISA} measurement errors and association with the LMC} \label{Sec:Parameter_Constraints}

\begin{figure*} 
    \centering
    \includegraphics[width=\textwidth]{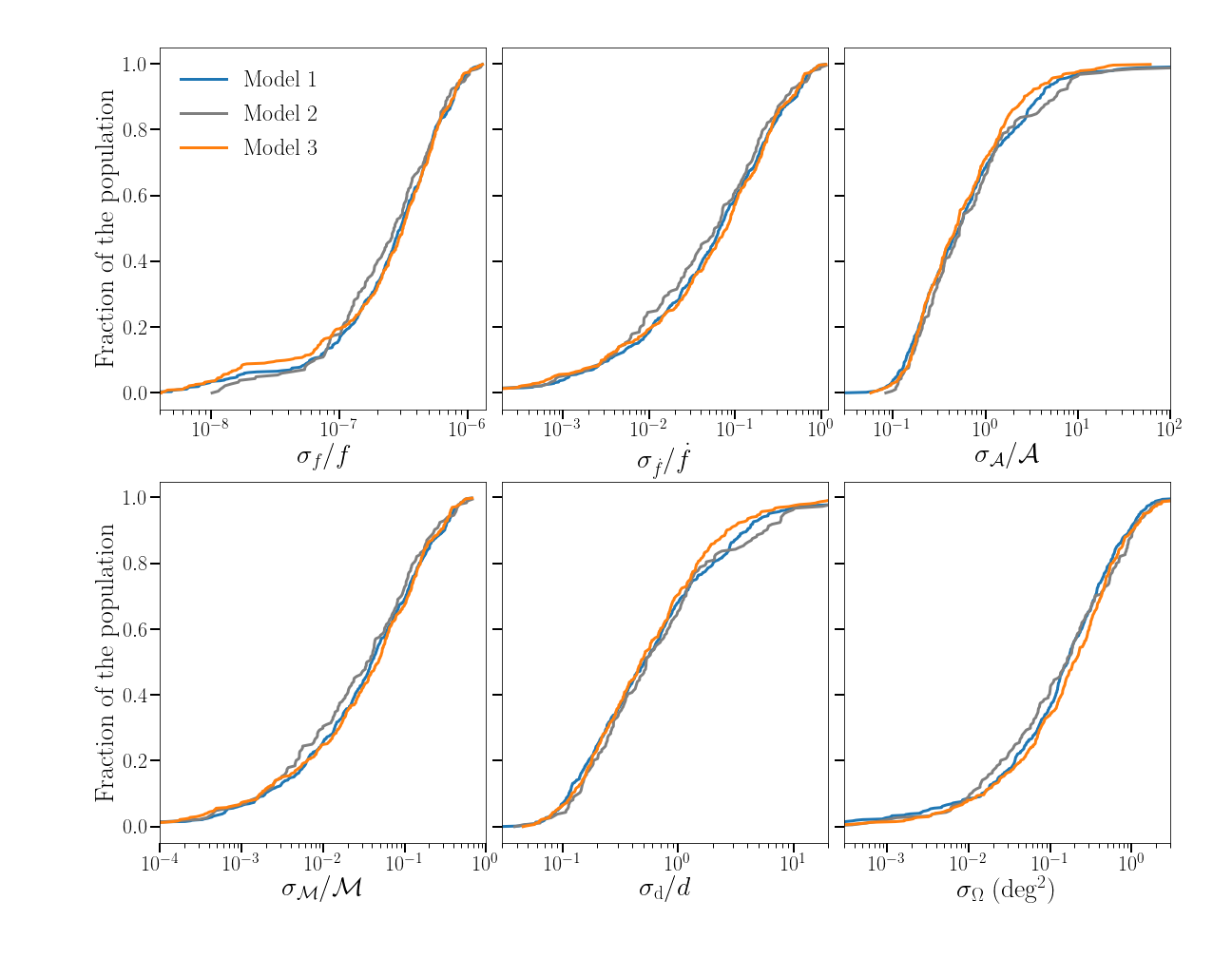}
    \caption{Cumulative distributions of errors for the {\it LISA} mission after 4\,yr. There errors are estimated with the Fisher information matrix technique as part of the {\it LISA} data analysis pipeline of \citet{kar21}.}\label{Fig:Cum_Err} 
\end{figure*}

\begin{figure*} 
    \centering
    \includegraphics[width=0.49\textwidth]{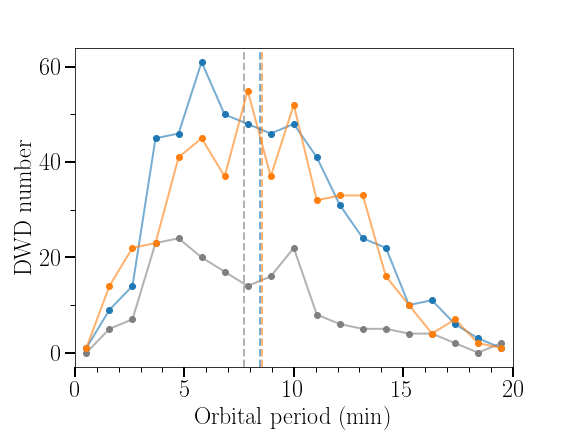}
    \includegraphics[width=0.49\textwidth]{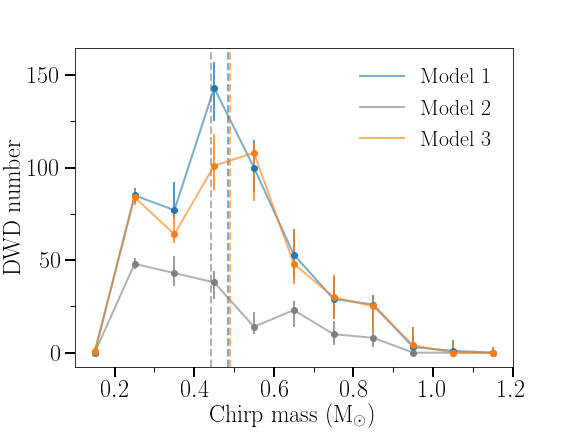}\quad

    \caption{Distributions of orbital periods (left) and chirp masses (right) of LMC DWD with S/N$>7$. As before we use blue colour for Model 1, grey for Model 2, and orange for Model 3. Error bars represent the minimum and maximum of each histogram bin in 100 realisations wherein periods (chirp masses) are drawn from a normal distribution based on $\sigma_{{\rm P},\:i}$ ($\sigma_{{\cal M},\:i}$) computed with FIM. Vertical dashed lines indicate the mean of the respective distributions. }\label{Fig:PErr} 
\end{figure*}

Fig.~\ref{Fig:Cum_Err} gives the cumulative distribution of estimated {\it LISA} errors using the FIM technique for frequency, frequency derivative, GW amplitude, chirp mass, distance and sky position. 
Since at the LMC distance {\it LISA} can only access DWDs with $f>$ a few mHz, frequency, frequency derivative and GW amplitude across the whole sample can be measured, which makes it possible to constrain binaries chirp masses and distances (cf. Eqs.~\ref{eqn:amp} and \ref{eqn:fdot}).
We also note that sky location is remarkably well constrained, with half of the sample having localisation errors of $\sigma_\Omega < 0.2$\,deg$^2$.  Using Bayesian techniques for the binary parameter estimation \citet{2020ApJ...894L..15R} found that high frequency DWDs in Milky Way satellites can have small sky localisation uncertainty, which improves exponentially with increasing GW frequency (or equivalently period). However, uncertainties in distance measurements have no such dependence. Indeed, half of the sample in Fig.~\ref{Fig:Cum_Err} have completely unconstrained distances because of the large errors on GW amplitude. This means that for the majority of LMC DWDs, error bars on the distance will be larger that the LMC physical size. 

Detected DWD binaries can be associated with the LMC because of the small errors on the sky position (< a few\,deg$^2$) compared to the angular size of the galaxy ($\sim$10\,deg) and because of its location in a sparsely populated part of the sky away from the Galactic plane. We quantify the probability of contamination by the (foreground) Milky Way DWDs by following \citet[][see their section 4]{2020ApJ...894L..15R}. They make use of the Milky Way (disc + bulge) DWD population from \citet{2019MNRAS.483.5518K} and include a stellar halo generated with a single burst SFH, a power law density distribution according to \citet{ior18}, and a total mass of $1.4\times10^9$\,M$_\odot$ \citep[e.g.][]{mak19}. They find that this Milky Way foreground model yields $\sim1$\,DWD/deg$^2$ in the sky region occupied by the LMC. Thus, with this foreground density and an average sky localisation error of 0.2\,deg$^2$, as estimated in this study, we obtain 0.2 contamination foreground sources or a typical false alarm probability of $\sim0.01$. We note that this result is frequency dependent: at lower (higher) frequencies, the sky localisation gets worse (better) and as a consequence the false alarm rate increases (decreases).
In addition, \citet{{2020ApJ...894L..15R}} point out that the situation may be further complicated by the fact that the LMC is located at the ecliptic pole, where {\it LISA} suffers by a partial degeneracy in the {\it LISA} response. This may result in a larger foreground due to the presence of Milky Way DWDs at the other ecliptic pole. Naturally, for binaries with precise distance estimates the associations will be robust.

\subsection{Parameter distributions of detectable DWDs}

In Fig.~\ref{Fig:PErr} we compare the distribution of detectable binary orbital periods and chirp masses for our LMC models. Error bars in Fig.~\ref{Fig:PErr} are estimated by taking the minimum and maximum of each histogram bin in 100 realisations, wherein each binary period/chirp mass are drawn from a normal distribution with a mean at the true period and a standard deviation set by the  period/chirp mass uncertainty (cf. Section~\ref{Sec:Parameter_Constraints}). We note that the error bars associated with the LMC DWD orbital period distribution (left panel) are very small because GW frequencies are very well constraint with $\sigma_{f}/f<10^{-6}$ for all binaries. From the comparison, it is clear that distributions of DWD orbital periods and chirp masses are similar in Model 1 and Model 3, while Model 2 differ from the other two. The similarities/differences can be attributed to similarity/differences in the star formation histories. Recall that Model 1 and 3 have similar star formation histories but different spatial distributions, while Model 2 has a different star formation history and shares the same spatial distribution as Model 3. In Model 2 the majority of DWD is formed within the first 4\,Gyr, whereas the majority of DWDs in Models 1 and 3 formed is the last 4\,Gyr (cf. Fig.\ref{Fig:SFR_eff}). In addition, $\sim$100 detectable binaries in Models 1 and 3 became DWD just a few Myr ago -- a truly `fresh' stock. This result reflects how period shortening through GW emission is a poor source of {\it LISA}-detectable binaries compared to the direct formation at low orbital period. In other words, it is more efficient to freshly populate the {\it LISA} band with newly born milli-Hz binaries than with old binaries that slowly shorten their period enough to be bright source in {\it LISA}. 

In Fig.~\ref{Fig:PErr} we also show the mean values of the respective distributions as dashed vertical lines, which fall at longer DWD orbital periods for Model 1 and Model 3 ($\sim$8.5\,min) and larger chirp masses ($\sim$0.49\,M$_\odot$) compared to Model 2 ($\sim$7.7\,min and $\sim$0.44\,M$_\odot$, respectively). In particular, we highlight that even a small difference of $\sim 0.7$\,min is well constrained by {\it LISA} considering that the largest relative error on the DWD orbital period across all three models is of 0.025\,ms. To give a more robust evidence that this lower relative period distribution is statistically meaningful, we apply a two sample Kolmogorov-Smirnov (KS) test which seeks to quantify the differences between two populations and to ascertain if they originated from the same probability distribution. We find that the difference between Model 1/Model 3 and Model 2 distribution is significant at the 0.005 level. Although, the chirp masses for the LMC DWDs will be overall well constrained (with the largest fractional error of 0.3), the average error is comparable to the difference between the means of the chirp mass distributions. Still, the KS test confirms that the chirp mass distribution are different at 0.001 significance level.

\subsection{Investigating constraints on the LMC stellar mass} \label{Sec:Mass_Constraints}

Fig.~\ref{Fig:results} clearly shows that the number of detectable DWDs is linearly dependent on the LMC's total stellar mass. Indeed, by performing a linear regression we find that our three models are well fit by
\begin{align}\label{Eq:Regression}
    N_{\rm M1} &\approx 195 \left( \frac{M_{\star}}{10^{9} M_\odot} \right)  \nonumber \\
    N_{\rm M2} &\approx 74.2 \left( \frac{M_{\star}}{10^{9} M_\odot} \right)  \\
    N_{\rm M3} &\approx 184 \left( \frac{M_{\star}}{10^{9} M_\odot} \right)  \nonumber 
\end{align}
where $N_{{\rm M}i}$ is the number of detectable DWD as a function of the LMC stellar mass $M_{\star}$. Note that the intercepts are approximately zero because zero LMC mass yields no DWDs. The coefficient of determination for each model takes on values that are reasonably close to 1 ($\sim$0.97, 0.75, and 0.97 for Models 1, 2, and 3, respectively) suggesting the relationship is indeed linear. We over-plot these results as dashed lines in Fig.~\ref{Fig:results}.

We can now exploit this linear dependence between the number of {\it LISA} detectable DWDs and the LMC mass, and reverse-engineer our modelling to infer what stellar mass is required to generate the `observed' number of DWD with {\it LISA}. Using as the study case of Milky Way satellite galaxies, \citet{kor21} showed that this can be done in analogy with simple stellar populating models widely used for estimating the stellar mass of galaxies based on their total light \citep[e.g.][]{tin76,bru83,mar98}. Using the same fiducial binary population synthesis model as in our work, they constructed a Bayesian inference scheme considering the number of detected (detached) DWDs associated with the satellite and the measured distance to the satellite as the only two input parameters. The authors demonstrated that counting the number of {\it LISA} sources in a satellite provides a measurement of its stellar mass. Even when the number of detections in a (known) satellite is zero, it is still possible to place an upper limit on the satellites mass using the same argument. We highlight that such an inference scheme recovers the `true' LMC mass and gives smaller errors when the star formation history used for the inferences coincides with the `true' one. Here we employ the same inference method as in \citealt{kor21} using a constant star formation history, which is different from star formation histories considered here to model the LMC. \citealt{2020A&A...638A.153K} predict that a constant star formation history for a total stellar mass of $2.7 \times10^9$\,M$_\odot$ would yield $\sim$235 {\it LISA} detectable DWDs. When comparing DWD formation histories in Fig.~\ref{Fig:SFR_eff} and our results in Table~\ref{Table:Mass}, not surprisingly this number falls in between that based on Model 2 (with no recent start formation) and that of Models 1 and 3 (with active star formation at recent times). We apply a Gaussian prior on DWD distances ($\mu=49.21$\,kpc and $\sigma=35\,$kpc), a flat prior for the LMC mass in the range $\left[10^6 - 10^{10}\right]\,$M$_\odot$ and a flat prior for the LMC age in the range $\left[1 - 10\right]\,$Gyr. We plug in the number of detected binaries estimated with Models 1, 2 and 3 for the true LMC mass of $2.7 \times 10^{9}\,$M$_\odot$ (cf. Table~\ref{Table:Mass}). 

\begin{figure} 
	\includegraphics[width=1\columnwidth]{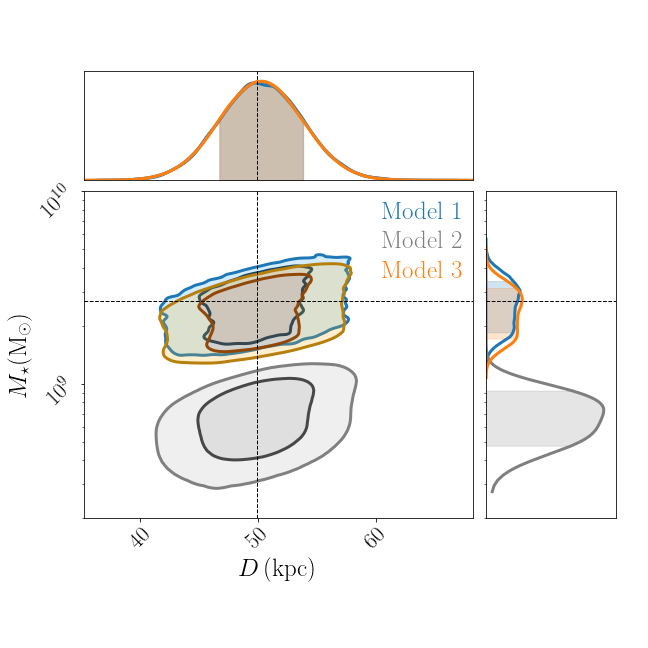}
    \caption{Posteriors for the LMC total stellar mass and distance for our three models: Model 1 (blue), Model 2 (grey), and Model 3 (orange). The assumed `true' LMC mass and distance are $2 \times 10^9\,$M$_\odot$ and $49.21\,$kpc (dashed black lines). We note that our Model 1 yields essentially the same contours as Model 3 - given that these models share the same star formation history.}
    \label{fig:LMCmass}
\end{figure}

We recover the LMC mass of $3.01^{+0.4}_{-1.1}\times 10^9\,$M$_\odot$ based on the number of detections generated by Model 1, $0.7^{+0.19}_{-0.25}\times 10^9\,$M$_\odot$ based on Model 2, and $2.7^{+0.39}_{-1.0} \times 10^9\,$M$_\odot$ based on Model 3. In Fig.~\ref{fig:LMCmass} we show the obtained posterior distribution for the LMC stellar mass, while the LMC age remains an unconstrained parameter. Fig.~\ref{fig:LMCmass} reveals that even using `incorrect' constant star formation history assumption, which is close enough to the `true' one as for the case of Model 1 and Model 3, we can recover the LMC mass within $1\sigma$.

\section{Discussion and conclusions} \label{Sec:Discussion}


In this work we assembled a realistic population of DWD binaries for a range of LMC stellar masses. We used a fiducial DWD synthetic model from \citet{2020A&A...638A.153K} based on binary population synthesis simulations of \citet{2012A&A...546A..70T} computed with the {\sc SeBa} code. These binaries were injected in three different LMC models conceived to test the effect of different star formation models and spatial distributions of stars on on the size and properties of {\it LISA} detectable DWD population. In Model 1 we assigned synthetic DWD binaries ages and positions based on the completeness-corrected regional star formation histories of \citet{2009AJ....138.1243H}. This model provides an observation based star formation history, but because it uses sky-projected 2D stellar density distribution it lacks small-scale LMC's distinct features. To account for this later deficiency and to probe whether {\it LISA} can spatially resolve the LMC structure, in Model 2 we employ the N-body hydrodynamical simulation of \citet[][see bottom panels of Fig.~\ref{Fig:Spatial_M1}]{2020Natur.585..203L}. In addition to the spatial distribution of the simulation particles we also make use of the their mass and ages, which provides us with an alternative (simulation driven) star formation history to that of Model 1. In Model 3 we combined the most realistic features of the previous two models: observation based star formation history from Model 1 and spatial distribution of stars from Model 2. For all three models we then compute how many DWD could be detected by {\it LISA} over a 4\,yr mission and how well we could constrain their parameters. Our main findings can be summarised as follows.

{\it LISA} will be able to detect from a hundred to several hundred DWDs in the LMC. These include both detached and accreting systems. In this study we do not provide a detailed modelling of the accreting DWD systems, thus we cannot confidently state if they establish a stable of mass transfer or merge. Even when we consider detached DWDs only, our most realistic LMC model (Model 3) yields from 159 to 357 {\it LISA} detections. This means that the LMC  will stand out as an extended and spatially resolved structure on the {\it LISA}'s GW sky.

Importantly, {\it LISA} will provide scores of double degenerate SNIa progenitors, extremely challenging to find with electromagnetic telescopes. Our estimates range from 25 to 125 detectable SNIa progenitors. To quote these numbers we arbitrary define as SNIa progenitor a DWD with ${\cal M}>0.6\,$M$_\odot$, which corresponds to an equal-mass binary exceeding the Chandrasekhar mass limit. Although our SNIa progenitor definition is oversimplified, it suggests that {\it LISA} will play a crucial role in understanding SNIa double-degenerate formation channel. 
SNIas are useful as a standard candles to measure cosmological distances~\citep{1998AJ....116.1009R,1999ApJ...517..565P}, and therefore a number of works have explored the role of DWDs in total SNIa merger rates~\citep{2012A&A...546A..70T,2012ApJ...749L..11B,2012ApJ...751..143M,2014A&A...563A..83C,2018MNRAS.476.2584M,2018A&A...610A..22T} as opposed to other SNIa formation channels such as those reviewed in~\citet{2018PhR...736....1L}. While identifying double degenerate SNIa progenitors can help in determining their contribution to the SNIa merger rate, no previously observed stellar system have been robustly confirmed to be such a  progenitor~\citep{2019MNRAS.482.3656R}. Our results show that {\it LISA} will be fundamental in discovering SNIa progenitors not only within our own Galaxy, but also in a completely different lower metallicity and actively star-forming environment as that of the LMC. 

We demonstrate that while the number of {\it LISA} detections is linearly dependent on the LMC total stellar mass, it is also sensitive to recent star formation activity (cf. Fig.~\ref{Fig:results}). This can be deduced when comparing Model 1 and Model 3 (with recent star formation) to Model 2 (with most of DWDs being formed at early times). We find that Model 1 and 3 generate 1.8 - 2.8 times more {\it LISA} detectable DWDs than Model 2. In particular, a large fraction of this difference comes from binaries formed $\sim\mathcal{O}(10^2)$ Myr ago, which are $\sim$10 times more numerous in Models 1 and 3 compared to Model 2 and became DWDs just in the last few Myr. To sum up, these results show that the total stellar mass sets the total number of DWDs in the LMC, while the star formation history (in combination with the age) determines how many detectable DWDs emit in the {\it LISA} frequency band at the present time.

Besides the total number of binaries, the difference is also noticeable when comparing DWD period and chirp mass distributions (cf. Fig.~\ref{Fig:PErr}). We show that DWDs in Model 1 and Model 3 on average have $\sim$1\,min shorter orbital periods and $\sim$0.05\,M$_\odot$ higher chirp masses  than DWDs in Model 2. This highlights how period shortening through GW emission is a less efficient way of forming {\it LISA} detectable binaries compared to the formation of binaries directly in the {\it LISA} band. The model comparison also reveals that the numbers of {\it LISA} detectable DWDs do not significantly change with the LMC spatial distribution. 
Further analysis may narrow down what can be learned about the LMC's morphology and characterise the GW confusion noise from unresolvable LMC DWDs. We note that the impacts of both stellar mass and the current star formation rate on the number of detectable binaries could be disentangled by considering the period distribution of detected binaries. One could imagine quantitatively relating the number of detectable binaries, the median period of detectable binaries, the stellar mass of the LMC, and the current star formation rate. However, such a relation is out of the scope of this work since our models only considered high star formation activity and near complete inactivity, and our approach includes only period distributions from a limited number of simulations. Nevertheless, our findings suggests that the period and chirp mass distributions and number of {\it LISA} detectable DWD binaries may probe both LMC stellar mass and recent star formation, motivating more careful study.

While we explore different masses and star formation histories, our methodology relies on a fiducial DWD population model. Though since the DWD formation we only consider energy loss through GWs, other processes such as tidal interactions may alter the size and the properties of the {\it LISA} sample \citep[e.g.][]{bis22}. As we have already mentioned, we do not provide detailed modelling for the accreting DWD, but only flag as accreting binaries that overfill their Roche Lobe at the time they are observed with {\it LISA}. It is yet unclear what is the most likely fate for these binaries, and so more development is needed from to both theory and observations to make reliable forecasts for {\it LISA} \citep[e.g.][]{LISAastroWP}. However, we chose to not remove these binaries from our simulations so that these can be followed up in the future studies. We also keep some of the key quantities entering in the binary evolution such as the metallicity, common envelope model and its efficiency, and binary fraction. Earlier work of \citet{2020A&A...638A.153K} showed these have minor effect compared to the assumptions on the properties of the satellite galaxy, which we extensively explore here.

Given the number and well constrained sky locations of detectable binaries, future structural analysis of the LMC using our population is highly promising. Additionally, while this work focuses on detectable binaries, far more will be undetectable and contribute to a background confusion noise. Future efforts may try to constrain this noise in case it is relevant for the  {\it LISA} sensitivity curve (as in \citealt{2017JPhCS.840a2024C}) or useful for probing structure (as in \citealt{2006ApJ...645..589B}). Similarly, the past observed and simulated studies our research depended on have also been conducted for the SMC \citep{2004AJ....127.1531H,2020Natur.585..203L} making similar analysis straightforward. Even so, the SMC is estimated to contain a factor of $\lesssim$5 fewer detectable DWD binaries compared to the LMC \citep{2020A&A...638A.153K} making the utility of gravitational probes less optimistic. Finally, even though the LMC is predicted to host only $\sim$2-5 detectable double neutron star binaries \citep{2020MNRAS.492.3061L,2019MNRAS.489.4513S}, their inclusion together with other types of compact object binaries in our work could only improve  {\it LISA}'s ability to extract LMC features. 


\section*{Acknowledgements} \label{Sec:Acknowledgements}

This work would not have been possible without Silvia Toonen, who provided us with a mock DWD population, and Scott Lucchini, who provided us with the LMC simulation. We also thank Nikolaos Karnesis for allowing us to use the data analysis pipeline outlined in Section~\ref{Sec:DA}. The code is publicly available in the following GitHub repository: \url{https://gitlab.in2p3.fr/Nikos/galactic_properties_from_cgbs}.\\
We also thank Jeffrey Pflueger, Fraser Evans, Stella Reino, and Pieter van Dokkum for useful conversations on implementing our methods and reporting results.\\
VK acknowledges support from the Netherlands Research Council NWO (Rubicon 019.183EN.015 grant). EMR acknowledges that this project has received funding from the European Research Council (ERC) under the European Union’s Horizon 2020 research and innovation programme (Grant agreement No. 101002511 - VEGA P)\\
This research made use of the open source python package PyGaia developer by Anthony Brown, publicly available at~\href{https://github.com/agabrown/PyGaia}{https://github.com/agabrown/PyGaia}, and tools provided by the \textit{LISA} Data Processing Group (LDPG) and the \textit{LISA} Consortium \textit{LISA} Data Challenges (LDC) working group~\href{https://lisa-ldc.lal.in2p3.fr/}{https://lisa-ldc.lal.in2p3.fr/}.


\section*{Data Availability} \label{Sec:Data_Availability}

Our fiducial LMC DWD population catalogue, which specifies the frequency, chirp, ecliptic latitude, ecliptic longitude, amplitude, inclination, polarisation, orbital phase, age, and white dwarf masses of present day {\it LISA} band LMC DWD binaries are available on Zenodo and can be accessed via \href{https://doi.org/10.5281/zenodo.6918083}{10.5281/zenodo.6918083}. This public release represents a 2.7$\times$10$^{9}$ M$_{\rm \odot}$ stellar mass LMC since we find the estimate of \citet{2002AJ....124.2639V} to be the most reliable of those considered here (being based on \citealt{1999AJ....118.2824Z}'s analysis of \citealt{1997AJ....114.1002Z} which is a more recent observational study than the work referenced in \citealt{1991rc3..book.....D} as was considered by \citealt{2012AJ....144....4M}). This public release excludes Model 2 since its star formation history does not reflect observation. Other population models are currently available upon request. We request that researchers utilising any of these populations cite this work.


\bibliographystyle{mnras}
\bibliography{biblio}

\appendix

\section{Model star formation rates} \label{sec: A}

\begin{figure}
    \centering
    \includegraphics[width=\columnwidth]{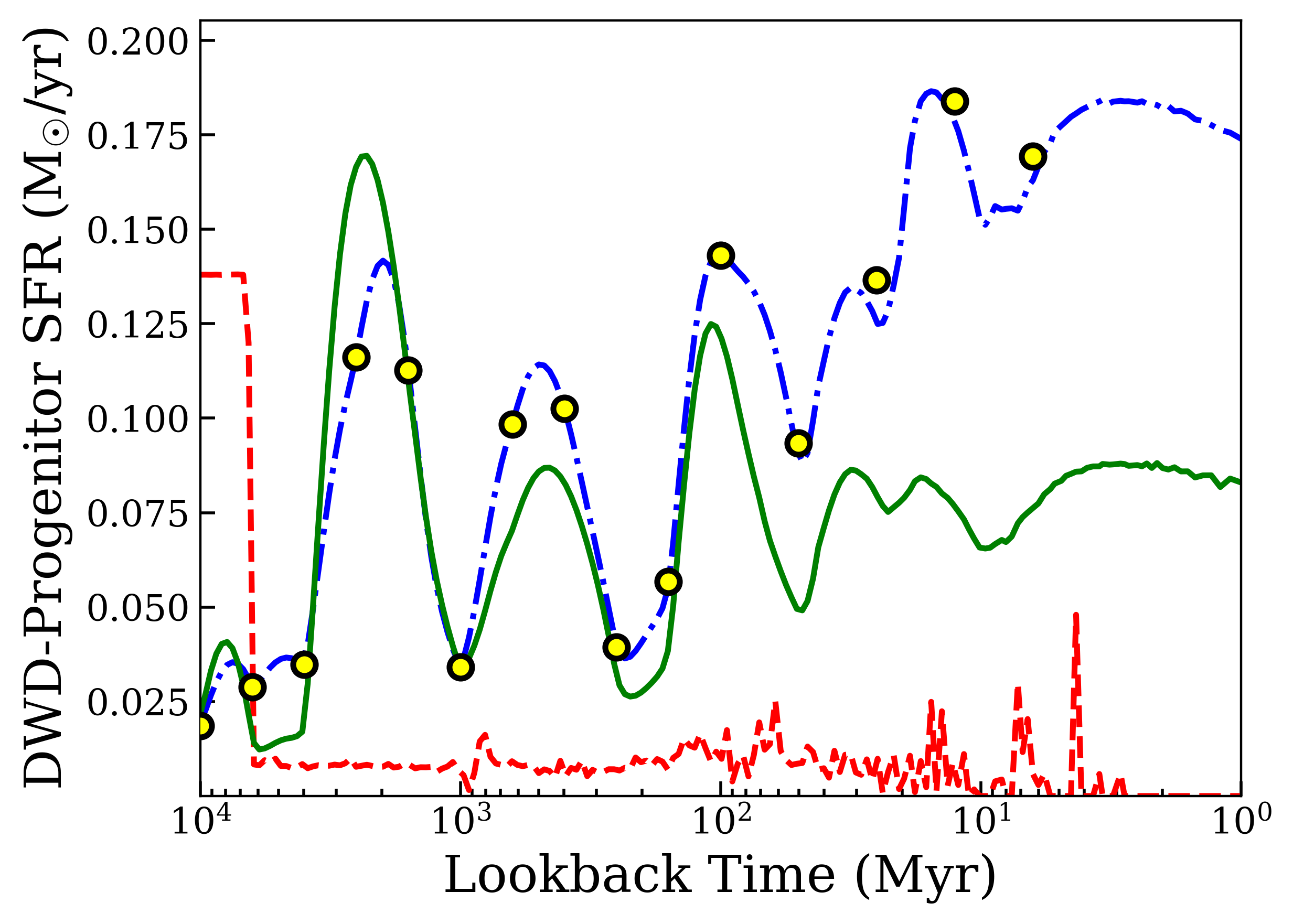}
    \caption{The total effective star formation history of MSMS binaries destined to become DWD binaries for Model 1 (the observation driven model; blue, dash-dotted), Model 2 (the simulated model; red, dashed), and Model 3 (the combined model; green, solid), along with the measurements of from~\citet[yellow, circles]{2009AJ....138.1243H} totalled across all regions and re-normalised by the global constant used in Model 1. While the $x$-axis is given in log scale, integrating over each star formation history gives the same MSMS population total mass. Note that this graph is given in logarithmic \textit{look-back} time as is probed by observation, rather than linear LMC time as given in Fig.~\ref{Fig:SFR_eff}.}\label{Fig:SFR_eff27} 
\end{figure}

In Fig.~\ref{Fig:SFR_eff27}, we show the total effective star formation history of each model assuming the LMC stellar mass of~\citet{2002AJ....124.2639V}. This is calculated by breaking up our simulation into logarithmically spaced time steps and dividing the mass of MS+MS binaries formed in each time step, totalled across the LMC, by the length of each time step. We see that Model 1's total star formation history agrees with the measurements of~\citet{2009AJ....138.1243H} once they are totalled across the LMC and re-normalised to reflect only the star formation rate of our DWD progenitors. Model 2 has little star formation in the last 6 Gyr, corresponding to the simulated stellar particle's large ages. Model 3's bursty star formation is highly reminiscent of Model 1, though slight differences occur since the matching procedure makes different LMC regions play a larger, or smaller, role in the total star formation history compared to Model 1.

\section{Number of {\it LISA} detections} \label{sec: B}

Table~\ref{Table:Mass} shows our results for the total number of binaries in the {\it LISA} band (N${^{\rm LISA}_{\rm DWD}}$) and binaries detected with S/N$>$7 (N${^d_{\rm DWD}}$). N${^d_{\rm DWD}}$ further beaks down into detached systems (N${^d_{\rm DDWD}}$), while the number of detectable accreting DWDs can be derived by subtracting these from the total. In the last column we highlight a sub-sample of {\it LISA}-detectable DWD that are potential double degenerate SNIa progenitors with ${\cal M}>0.6$\,M$_{\odot}$.
We predict a few to several hundreds of individually resolved DWD binaries in the LMC, that correspond to a detection efficiency in the {\it LISA} band of  $N{^d_{\rm DWD}}/N{^{\rm LISA}_{\rm DWD}} \sim 10^{-4}$. Therefore, for each resolved binary there are $10^{4}$ low signal-to-noise ones that can contribute to a --spatially localised --background signal. The detected ones are between 5\,per cent-10\,per cent of the N${^{<20{\rm min}}_{\rm DWD}}$ populations in the {\it LISA} band.

\begin{table*}
\caption{The impact of stellar mass and our different models for age and position assignment on the population of DWD binaries with a {\it LISA} S/N$>$7 after the nominal 4 yr mission length.}
\label{Table:Mass}      
        \begin{tabular}{cccccc}
        \hline \noalign{\vskip 0.1em}
        \multicolumn{1}{c}{Mass (M$_{\rm \odot}$)} & \multicolumn{1}{c}{Model} &
        \multicolumn{1}{c}{N${^{\rm LISA}_{\rm DWD}}$} &
        \multicolumn{1}{c}{N${^d_{\rm DWD}}$} & \multicolumn{1}{c}{N${^d_{\rm DDWD}}$} & \multicolumn{1}{c}{N${^d_{\rm SNIa}}$} \\
        \noalign{\vskip 0.1em}
        \hline 
1.5$\times$10$^{9}$ & 1 & 1.21$\times$10$^{6}$ & 257 & 161 &  71 \\ 
1.5$\times$10$^{9}$ & 2 & 1.32$\times$10$^{6}$ & 119 &  70 &  25 \\ 
1.5$\times$10$^{9}$ & 3 & 1.24$\times$10$^{6}$ & 258 & 159 &  68 \\ \hline 
2.0$\times$10$^{9}$ & 1 & 1.61$\times$10$^{6}$ & 333 & 213 &  90 \\ 
2.0$\times$10$^{9}$ & 2 & 1.76$\times$10$^{6}$ & 184 &  92 &  39 \\
2.0$\times$10$^{9}$ & 3 & 1.65$\times$10$^{6}$ & 348 & 218 &  77 \\ \hline
2.5$\times$10$^{9}$ & 1 & 2.01$\times$10$^{6}$ & 483 & 303 &  93 \\ 
2.5$\times$10$^{9}$ & 2 & 2.20$\times$10$^{6}$ & 201 & 105 &  39 \\ 
2.5$\times$10$^{9}$ & 3 & 2.06$\times$10$^{6}$ & 431 & 282 & 111 \\ \hline
2.7$\times$10$^{9}$ & 1 & 2.17$\times$10$^{6}$ & 517 & 323 & 112 \\
2.7$\times$10$^{9}$ & 2 & 2.37$\times$10$^{6}$ & 184 &  83 &  41 \\ 
2.7$\times$10$^{9}$ & 3 & 2.23$\times$10$^{6}$ & 465 & 294 & 107 \\ \hline
3.0$\times$10$^{9}$ & 1 & 2.42$\times$10$^{6}$ & 517 & 318 & 111 \\
3.0$\times$10$^{9}$ & 2 & 2.64$\times$10$^{6}$ & 256 & 128 &  39 \\
3.0$\times$10$^{9}$ & 3 & 2.47$\times$10$^{6}$ & 543 & 357 & 125 \\ \hline
\end{tabular} \\
\contcaption{Column 1: total LMC stellar mass. 2: distribution model (1 $\equiv$ observation driven, 2 $\equiv$ simulation driven, 3 $\equiv$ combined). 3: number of of present day DWD binaries in the {\it LISA}-band with $f>10^{-4}$ Hz. 4: number of {\it LISA}-detectable DWD binaries as calculated from the MLDC pipeline. 5: number of {\it LISA}-detectable DWD binaries that are detached, engaging in negligible mass transfer. 6: number of {\it LISA} detectable DWDs that are potential double degenerate SNIa progenitors with ${\cal M}>0.6$\,M$_{\odot}$.}
\end{table*}

\label{lastpage}
\end{document}